\documentclass[3p,times,twocolumn]{elsarticle}

\usepackage{ecrc}

\newcommand{\be}{\,\begin{equation}}
\newcommand{\ee}{\,\end{equation}}

\volume{00}

\firstpage{1}

\journalname{Nuclear Physics B Proceedings Supplement}

\runauth{}

\jid{nuphbp}

\jnltitlelogo{Nuclear Physics B Proceedings Supplement}

\usepackage{amssymb}
\usepackage[figuresright]{rotating}

\begin{document}

\begin{frontmatter}



\dochead{}

\title{Recent developments in cosmic ray physics}


\author{P. Blasi}

\address{INAF/Osservatorio Astrofisico di Arcetri, Largo E. Fermi, 5 50125 Firenze, Italy \\
Gran Sasso Science Institute (INFN), Viale F. Crispi 6, 60100 L'Aquila, Italy}

\begin{abstract}
The search for a theory of the origin of cosmic rays that may be considered as a standard, agreeable model is still ongoing. On one hand, much circumstantial evidence exists of the fact that supernovae in our Galaxy play a crucial role in producing the bulk of cosmic rays observed on Earth. On the other hand, important questions about their ability to accelerate particles up to the knee remain unanswered. The common interpretation of the knee as a feature coinciding with the maximum energy of the light component of cosmic rays and a transition to a gradually heavier mass composition is mainly based on KASCADE results. Some recent data appear to question this finding: YAC1 - Tibet Array and ARGO-YBJ find a flux reduction in the light component at $\sim 700$ TeV, appreciably below the knee. Whether the maximum energy of light nuclei is as high as $3000$ TeV or rather as low as a few hundred TeV has very important consequences on the supernova remnant paradigm for the origin of cosmic rays, as well on the crucial issue of the transition from Galactic to extragalactic cosmic rays. In such a complex phenomenological situation, it is important to have a clear picture of what is really known and what is not. Here I will discuss some solid and less solid aspects of the theory (or theories) for the origin of cosmic rays and the implications for future searches in this field.  
\end{abstract}


\end{frontmatter}

% \linenumbers

\section{Introduction}
\label{sec:intro}

In the long search for the origin of cosmic rays (CRs), the current period appears particularly rich with new data as well as with both old and new unanswered questions. Larger, better experiments allowed us to collect an impressively large wealth of data on spectra, mass composition and, although to a lesser extent, anisotropy, so much so that the main limitation to using such data is imposed in most cases by systematic errors and systematic uncertainties rather than statistics. The spectral breaks measured by PAMELA \cite{Adriani:2011cu} in the proton and helium spectra and not confirmed by the preliminary results of AMS-02, presented at the 33$^{rd}$ ICRC, might be instances of this situation. 

For many years, one of the main observational pillars in the search for the origin of CRs has been the measurement of a knee in the light cosmic ray component at about the same energy as the knee in the all-particle spectrum \cite{Antoni:2005wq}. This finding has driven much of the recent theoretical investigation on the highest energies achievable in supernova remnants, in that reaching the knee, even in the presence of rapidly growing plasma instabilities, appears to be rather problematic. As a confirmation on how potentially severe systematic uncertainties may affect our findings, YAC1 - Tibet Array \footnote{Results presented at the $33^{rd}$ ICRC in Rio de Janeiro, Brazil, 2013.}, and independently ARGO-YBJ \footnote{Results presented by I. De Mitri at Frontier Research in Astrophysics, May 26-31, 2014, Mondello, Italy} have measured the spectrum of the light component in CRs, and claimed that a knee may be identified at energy $\sim 700$ TeV, appreciably below the all-particle spectrum knee. ARGO-YBJ also managed to measure the all-particle spectrum, confirming the detection of the standard knee at $\sim 3\times 10^{15}$ eV. Interestingly, we are still rather uncertain as to whether the proton spectrum steepens at $\sim 400$ TeV or $\sim 3000$ TeV. The implications of this data-driven ambiguity on both the modelling of acceleration in Galactic sources and on the end of the Galactic CR spectrum and transition to extragalactic CRs are rather severe. 

From the theoretical point of view, the basic model for the origin of Galactic cosmic rays is based on two pillars: {\it a)} CRs are assumed to be accelerated with power law spectra, $N(E)\sim E^{-\gamma}$ in Galactic sources, such as supernova remnants (SNRs). {\it b)} CRs propagate diffusively throughout the Galaxy with a diffusion coefficient that, at energies above a few GeV, is assumed to be $D(E)\sim E^{\delta}$. The basic combination of these phenomena, acceleration and propagation in the Galaxy, leads to an equilibrium spectrum $n(E)\sim N(E)/D(E)\propto E^{-(\gamma+\delta)}$, that reflects the balance between injection of newly accelerated CRs and escape of CRs from the confinement volume on time scales $\tau_{d}=H^{2}/D(E)$, where $H$ is the height of the galactic halo. This simple estimate also leads to the powerful prediction that the ratio of fluxes of secondary and primary nuclei should be a decreasing function of energy, $\sim E^{-\delta}$ at high energy, so that the measurement of such ratios (such as the Boron to Carbon ratio) would allow us to measure the energy dependence of the diffusion coefficient. 

Both pillars listed above, acceleration and propagation, acquire a physical meaning only after a derivation from some basic physical principles. When one tries to do so, it is easy to show that several subtleties can change the simple predictions discussed above in ways that are potentially subject to observational scrutiny. Here I will briefly discuss several of these effects, while a more complete discussion can be found in recent reviews such as those in Refs. \cite{2013A&ARv..21...70B,2013arXiv1312.1590B,2014arXiv1406.7714A}.

The paper is organized as follows: in \S \ref{sec:prop} I discuss some simple modifications of the standard diffusion scenario that lead to important changes in predicted CR spectra and secondary to primary ratios. In \S \ref{sec:accel} I introduce some of the most important non-linear effects that are at the very basis of the mechanism of diffusive shock acceleration in SNRs. In \S \ref{sec:trans} I comment on the implications of the SNR paradigm for the end of the Galactic CR spectrum and the transition to extragalactic CRs. I summarize in \S \ref{sec:summary}.

\section{CR transport in the Galaxy}
\label{sec:prop}

Current models of CR propagation mainly focus on two scenarios: 1) {\it pure diffusion model}, with $D(E)\propto E^{0.6}$ and a flattening below a few GeV, and injection spectrum $N(E)\sim E^{-2.1}$; 2) {\it reacceleration model}, with $D(E)\propto E^{1/3}$ and no low energy break, and an injection spectrum $N(E)\propto E^{-2.4}$ with a low energy break below $\sim 2$ GeV. 
In fact there is a third class of propagation models that postulate the presence of a Galactic wind that leads to advection dominated CR transport at low enough energies, but the implications of these models for the high energy behaviour of the relevant quantities is not very different from that of the two classes of models listed above. 
In all these models the breaks in either the diffusion coefficient and/or the injection spectrum are required in order to fit the data and do not derive from fundamental physical arguments. Moreover, both pure diffusion models and reacceleration models are based on the assumptions of spatially homogeneous and isotropic diffusion. 

In the following I will discuss the implications of relaxing some of these assumptions in a minimal way. 

\subsection{Anisotropic Diffusion}

In the presence of a large scale magnetic field, such as the spiral shaped one in the Galaxy, diffusion parallel to the large scale field is bound to be faster than diffusion perpendicular to the field lines. This is true even in the presence of fluctuations with $\delta B/B\sim 1$, as discussed for instance in \cite{2002PhRvD..65b3002C,2007JCAP...06..027D}. Observations of diffuse backgrounds of radio and gamma ray emissions require a significant diffusion of CRs perpendicular to the Galactic disc, which reflects in the need to have substantial random walk of magnetic field lines. In all existing models of CR propagation in the Galaxy this very important effect is not taken into account, although some recent calculations \cite{2014NewA...30...32K} have renewed the interest in this line of research.

\subsection{Inhomogeneous diffusion coefficient}

Even retaining the approximation of isotropic diffusion, the minimal generalization of the diffusion model consists in assuming that the diffusion coefficient may be larger in the halo of the Galaxy and smaller close to the Galactic disc, and possibly with different slopes, $E^{\delta_{1}}$ close to the disc and $E^{\delta_{2}}$ ($\delta_{2}>\delta_{1}$) farther away. This simple generalization of the diffusion model leads to equilibrium spectra of CRs that are no longer pure power laws, but broken power laws. The slope of the CR spectrum measured in the disc is $E^{-(\gamma+\delta_{2})}$ at low energies and $E^{-(\gamma+\delta_{1})}$ at high energy, where the boundary between low and high energies is set at some critical value $E_{*}$ determined by the absolute normalizations of the diffusion coefficients and by the spatial size of the regions 1 and 2 with different diffusion properties. This type of two zones diffusion has been recently studied \cite{2012ApJ...752L..13T} in connection with the detection of a break in the proton and helium spectra as measured by the PAMELA experiment \cite{Adriani:2011cu}. 

\subsection{CR induced diffusion} 

It has been known since the early '70s that the gradient of CRs diffusing in the Galaxy leads to the development of streaming instability and the generation of Alfv\'{e}n waves moving in the direction of the decreasing CR density \cite{1975Natur.258..687S,1975MNRAS.173..255S,1974MNRAS.166..155H,1980ARA&A..18..289C}. These waves can resonate with CRs thereby enhancing their scattering frequency. This causes a sort of self-confinement of CRs in the Galaxy, although the mechanism becomes increasingly less effective with increasing energy. The instability is assumed to saturate because of the competing effect of non-linear Landau damping. In \S \ref{sec:ion-neutral} we will comment on the role of ion-neutral damping. 

In the presence of a background turbulence, that may be due to injection in supernova explosions on large spatial scales, $\lambda\sim 100 pc$ and cascading of this power on smaller scales, it was recently shown \cite{Blasi:2012yr,Aloisio:2013tda} that the CR spectrum observed at the Earth may develop spectral breaks quite reminiscent of the ones observed by PAMELA \cite{Adriani:2011cu}. The B/C ratio and similar secondary to primary ratios can also be well described within this class of models \cite{Aloisio:2013tda}. 

The importance of the self-generation of turbulence by CRs also finds its justification in the fact that the predicted diffusion coefficient has an energy dependence $D(E)\propto E^{0.7}$, not too dissimilar from the standard diffusion models where the diffusion coefficient is assumed a priori to fit the data. In Ref. \cite{1997A&A...321..434P} the effect of self-generation was also combined with the launching of a Galactic wind induced by the CR pressure gradient. In this case the energy dependence of the diffusion coefficient was found to be somewhat shallower, $D(E)\propto E^{0.54}$. In both cases the expected anisotropy is larger than observed, although, as discussed in Ref. \cite{2005IJMPA..20.6858Z}, the anisotropy at the Earth might be affected by local phenomena such as a reduction of the diffusion coefficient in the neighbourhood of the solar system. 

\subsection{Ion-neutral damping}
\label{sec:ion-neutral}

An Afv\'{e}n wave propagating in a partially ionized medium gets damped with a damping rate $\Gamma_{D}=\nu/2$ \cite{1971ApL.....8..189K}, where $\nu=8.4\times 10^{-9} n_{1} T_{4}^{0.4}~\rm s^{-1}$ is the rate of scattering of neutrals in a gas with density $n_{1}=n/1 cm^{-3}$ and temperature $T_{4}=T/10^{4}$ K. This expression for the damping is valid for wavenumbers $k>k_{*}$, where $k_{*}=(\nu/v_{A})(1+n_{i}/n_{H})$. At $k<k_{*}$ the damping rate decreases proportional to $k^{2}$. The damping time scale for $k<k_{*}$ is of order a few tens of years for typical parameters of the interstellar medium (ISM). This corresponds to particles with momentum $p<p_{*}$ where $p_{*}\sim 75$ GeV/c is calculated by requiring the resonance condition for particles with Larmor radius $r_{L}(p_{*})$, $k_{*}\approx 1/r_{L}(p_{*})$. In other words, for the bulk of CRs in the Galaxy the ion-neutral damping is so fast that there should be very small scattering with waves, and particles should basically stream freely. Yet, we know that CRs diffuse (on average) in our Galaxy. This might mean that most of the diffusion volume filled by CRs is made of ionized gas, while neutrals are spatially segregated in molecular clouds, or might point at some flaw in our ideas on CR diffusion in the Galaxy. Regretfully, the problem of ion-neutral damping was not granted much attention in recent times. 

\section{CR acceleration in SNRs}
\label{sec:accel}

The SNR paradigm is based on the requirement that CRs may be accelerated in SNRs with an efficiency of the order of $\sim 10\%$ through diffusive shock acceleration (DSA) up to proton energy of the order of $\sim 3\times 10^{15}$ eV ($Z$ times larger for nuclei of charge $Z$). Both the requirements in terms of efficiency and maximum energy demand that non-linear effects play an important role, as we discuss below. 

\subsection{Non-linear effects on the spectrum}

In the context of the test particle theory of particle acceleration at a strong shock, the spectrum of accelerated particles is $Q(E)\sim E^{-2}$ (in terms of the distribution function in momentum the injection would be $Q(p) \sim p^{-4}$); since the CR spectrum observed at the Earth has a slope $\sim 2.7$, the rule $\gamma+\delta=2.7$ implies that the required diffusion coefficient should be $D(E)\propto E^{0.7}$. The process of diffusive particle acceleration at a shock when the dynamical reaction of the accelerated particles is taken into account is described by the non-linear theory of DSA (NLDSA) (see \cite{2001RPPh...64..429M} for a review). The main predictions of NLDSA are the following: 1) the spectrum of accelerated particles is no longer a power law, and becomes concave, namely steeper than $p^{-4}$ at low energies ($p\lesssim 10 m_{p}c$) and harder than $p^{-4}$ at high energies (namely harder than $E^{-2}$ in terms of kinetic energy). 2) Conservation of energy and momentum across the shock, including accelerated particles, implies that since part of the ram pressure of the upstream plasma gets converted into accelerated particles, there is less energy available for conversion into thermal energy of the background plasma, therefore the temperature of the downstream gas is lower than expected in the absence of accelerated particles. 3) The super-alfvenic drift of accelerated particles with the shock results in several CR-induced plasma instabilities, that may facilitate the process of particle acceleration by creating the scattering centers necessary to shorten the acceleration time and reach higher energies. In the absence of this phenomenon it is easy to show that the maximum energy achievable in a SNR is in the GeV range, rather than close to the knee, as observations require. 

\begin{figure}
\includegraphics[width=0.4\textwidth]{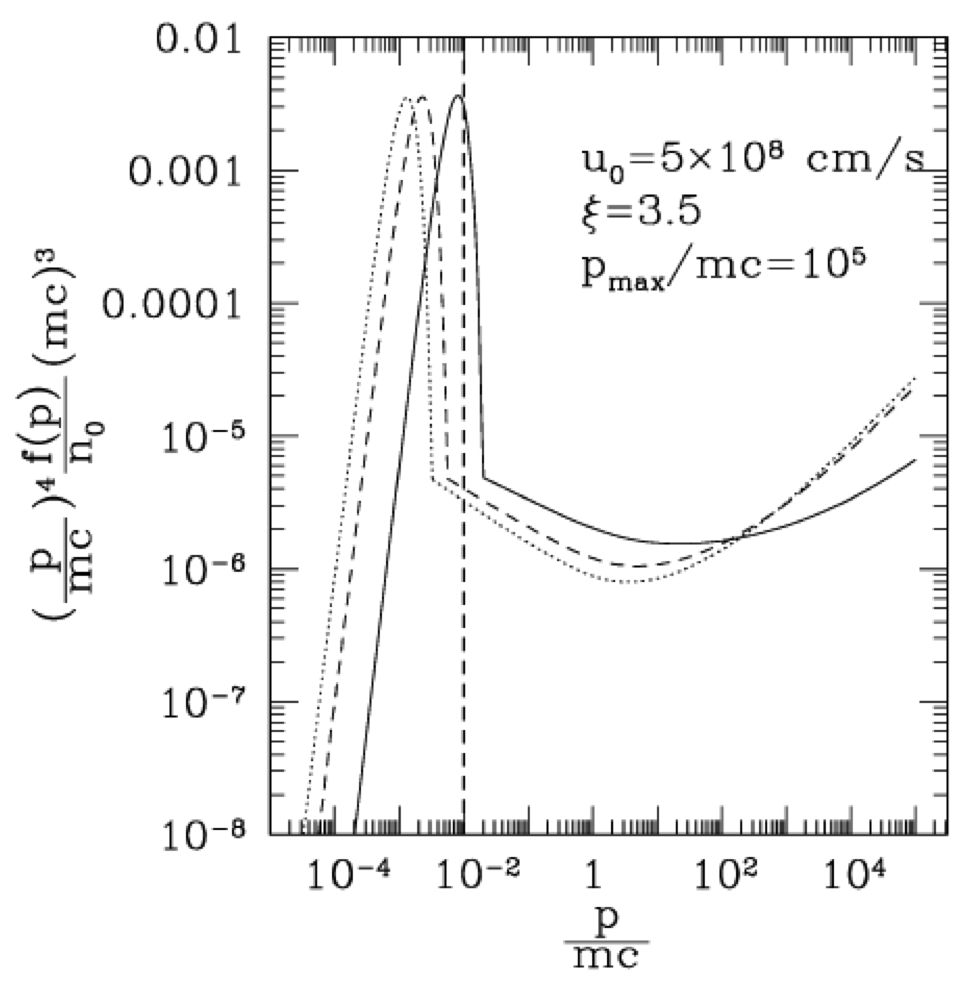}
\caption{Particle spectra (thermal plus non-thermal) at a CR modified shock with Mach number $M_{0}=10$ (solid line), $M_{0}=50$ (dashed line) and $M_{0}=100$ (dotted line). The vertical dashed line is the location of the thermal peak as expected for an ordinary shock with no particle acceleration (this value depends very weakly on the Mach number, for strong shocks). The plasma velocity at upstream infinity is $u_{0}=5\times 10^{8}$ $cm/s$, $p_{max}=10^{5} m_{p}c$ and the injection parameter is $\xi=3.5$ \cite{2005MNRAS.361..907B}.}
\label{fig:modshockspec}      
\end{figure}

The first two effects are illustrated well in Fig. \ref{fig:modshockspec} (from \cite{2005MNRAS.361..907B}) where I show the spectrum (thermal plus non-thermal) of the particles in the shock region for a shock Mach number $M_{0} = 10$ (solid line), $M_{0} = 50$ (dashed line) and $M_{0} = 100$ (dotted line). The thermal component is assumed to have a Maxwellian shape. Increasing the Mach number of the shock the acceleration efficiency is shown to increase, so that the spectra become increasingly more concave (the dynamical reaction of accelerated particles increases). At the same time, the peak of the thermal distribution moves leftward, namely the background plasma becomes colder for larger CR acceleration efficiency. The vertical dashed line in Fig. \ref{fig:modshockspec} shows the position of the thermal peak for a shock that does not accelerate CRs. 

One can appreciate that the spectrum of accelerated particles at high momenta becomes harder than $p^{-4}$ (namely harder than $E^{-2}$). It follows that the value of $\delta$ required to fit observations when NLDSA is used is even larger than 0.7. Realistic calculations \cite{2007ApJ...661L.175B} find $\delta\simeq 0.75$. 

In both cases of DSA and NLDSA the injection spectrum is expected to be at least as hard as $E^{-2}$ at high energy. The correspondingly large value of $\delta$ implies an exceedingly large anisotropy \cite{Ptuskin:2006p666,Blasi:2012p2024} which suggests that probably our descriptions of either particle acceleration or anisotropy are unsatisfactory. In this respect it is worth recalling the interesting exercise in Ref. \cite{2005IJMPA..20.6858Z}, where it was shown that a reduction of the diffusion coefficient in the solar neighborhood would lead to a reduction of the expected anisotropy while leaving the CR fluxes basically unaffected. 

In the last few years several SNRs have at last been detected in gamma rays, often from the GeV to the TeV energy range. This gamma ray emission is the result of inverse Compton scattering of electrons and pion production in inelastic hadronic collisions. The latter process has been viewed for a long time as the smoking gun of CR acceleration in SNRs. Ref. \cite{Caprioli:2011p2134} pointed out that the CR spectra inferred from gamma ray observations of a sample of SNRs are substantially steeper than the prediction of NLDSA (and of DSA as well). This finding suggests that the problem discussed above might be in our understanding of the process of particle acceleration. Indeed, we are aware of several effects that might cause the appearance of steeper spectra of accelerated particles, although all of them are rather speculative at present. 

If the velocity of the scattering centers \cite{Ptuskin:2010p1025,Caprioli:2010p133} responsible for particles' diffusion around the shock surface is large enough, the spectra of accelerated particles can become appreciably steeper than $E^{-2}$, since the relevant compression factor is not the ratio of the upstream and downstream fluid velocities, but rather the ratio of the upstream and downstream velocity of the scattering centers, as seen in the shock frame. This effect depends on the helicity of the waves responsible for the scattering, and may in principle lead to a hardening of the spectrum rather than a steepening. In turn the helicity of the waves depends on numerous aspects of wave production which are very poorly known and are hardly accessible observationally. 

A very important point to keep in mind is that the spectrum of accelerated particles as calculated at any given time at a SNR shock is not the same as the spectrum of particles leaving the remnant to become CRs. The process that connects the accelerated particles with CRs is the escape of particles from the accelerator \cite{2011MNRAS.415.1807D}. The understanding of this process is one of the most challenging aspects of the SNR paradigm. During the Sedov-Taylor phase of the expansion of a SNR, particle escape can only occur at momenta close to the maximum momentum reached at that given time, therefore the instantaneous spectrum of escaping particles is peaked around $p_{max}(t)$. On the other hand, this escape flux becomes close to a power law $\sim E^{-2}$ once it is integrated in time \cite{Caprioli:2010p133,Ptuskin:2010p1025}. Non-linear effects make the spectrum of escaping particles somewhat harder than $E^{-2}$. The accelerated particles that are advected downstream of the shock lose energy adiabatically and can leave the remnant only when the SNR shock disappears. The total escape flux from a SNR was calculated in the context of NLDSA \cite{Caprioli:2010p133,Ptuskin:2010p1025}, where it was shown that the spectrum of escaping particles is somewhat steeper than $E^{-2}$ and less concave than the instantaneous spectrum at any given time.  

Since the CR spectrum inferred from gamma ray observations is very close to the instantaneous spectrum of accelerated particles, the above considerations on escape do not affect the problem of the comparison of NLDSA with gamma ray observations \cite{Caprioli:2011p2134}.

\subsection{Magnetic field amplification and maximum energy}

As mentioned above, arguably the most crucial aspect of NLDSA is the phenomenon of magnetic field amplification. There are two aspects of this problem: 1) magnetic field amplification is required in order to explain the thin non-thermal X-ray rims observed in virtually all young SNRs; 2) the phenomenon is also likely to be related with the need for enhanced scattering of particles close to the shock surface. 

Magnetic field amplification can occur downstream due to plasma related phenomena \cite{Giacalone:2007p962} if the shock propagates in an inhomogeneous medium with density fluctuations $\delta\rho/\rho\sim 1$. While crossing the shock surface these inhomogeneities lead to shock corrugation and to the development of eddies in which magnetic field is frozen. The twisting of the eddies may lead to magnetic field amplification on time scales $\sim L_{c}/u_{2}$, where $L_{c}$ is the spatial size of these regions with larger density and $u_{2}$ is the plasma speed downstream of the shock. Smaller scales also grow so as to form a power spectrum downstream. This phenomenon could well be able to account for the observed thin X-ray filaments. The acceleration time for particles at the shock is however not necessarily appreciably reduced in that no field amplification occurs upstream of the shock. Hence, this mechanism may be important for particle acceleration at perpendicular shocks. It seems unlikely however that this scenario, so strongly dependent upon the geometry of the system, may lead to a general solution of how to reach the highest energies in Galactic CRs, although this possibility definitely deserves more attention. 

A more interesting possibility, that may address at the same time the problem of magnetic field amplification and that of making particle acceleration faster, is related to CR induced magnetic field amplification. 

It has been known for quite some time that the super-Alfv\'enic streaming of charged particles in a plasma leads to the excitation of an instability \cite{Skilling:1975p2165}. The role of this instability in the process of particle acceleration in SNR shocks was recognized and its implications were discussed by many authors, most notably \cite{Zweibel:1979p2308} and \cite{Achterberg:1983p2080}. The initial investigation of this instability led to identify as crucial the growth of resonant waves with wavenumber $k=1/r_{L}$, where $r_{L}$ is the Larmor radius of the particles generating the instability. The resonance condition, taken at face value, leads to expect that the growth stops when the turbulent magnetic field becomes of the same order as the pre-existing ordered magnetic field $\delta B\sim B_{0}$, so that the saturation level of this instability has often been assumed to occur when $\delta B/B\sim 1$. Refs. \cite{Lagage:1983p1347,Lagage:1983p1348} used this fact to conclude that the maximum energy that can possibly be reached in SNRs when the accelerated particles generate their own scattering centers is $\lesssim 10^{4}-10^{5}$ GeV/n, well below the energy of the knee. Hence, though the streaming instability leads to an appealing self-generation of the waves responsible for particle diffusion, the intrinsic resonant nature of the instability would inhibit the possibility to reach sufficiently high energy. 

This simple conclusion requires however some further analysis: for simplicity let us consider the case of a spectrum of accelerated particles coincident with the canonical DSA spectrum $f_{CR,0}(p)\propto p^{-4}$ for $\gamma_{min} \leq p/m_{p}c \leq \gamma_{max}$. When the CR efficiency is small, namely when the condition
\be
\frac{n_{CR}}{n_{i}} \ll \frac{v_{A}^{2}}{V_{sh}c}
\label{eq:condition}
\ee
is fulfilled \cite{Zweibel:1979p2308,Achterberg:1983p2080}, it is easy to show that Alfv\'en waves are excited (namely $Re\left[\omega\right]\approx k v_{A}$) and their growth rate is:
\be
Im\left[\omega\right](k)  \equiv \omega_{I} (k)= \frac{\pi}{8} \Omega_{p}^{*} \frac{V_{sh}}{v_{A}}\frac{n_{CR}(p>p_{res}(k))}{n_{i}}.
\label{eq:Im}
\ee
This is exactly the case discussed above. 

As an order of magnitude the density of CRs can be related to the efficiency of CR acceleration as $\frac{n_{CR}}{n_{i}} \approx \frac{3\xi_{CR}}{\gamma_{min}\Lambda}\left( \frac{V_{sh}}{c}\right)^{2}$,
where $\gamma_{min}\sim 1$ is the minimum Lorentz factor of accelerated particles and $\Lambda=\ln(\gamma_{max}/\gamma_{min})$, $\gamma_{max}$ being the maximum Lorentz factor. Eq. \ref{eq:condition} becomes then
\be
\xi_{CR} \ll \frac{\gamma_{min}\Lambda}{3} \left( \frac{v_{A}}{V_{sh}}\right)^{2}\frac{c}{V_{sh}}\approx 8\times 10^{-4} \left( \frac{V_{sh}}{5\times 10^{8} cm/s}\right)^{-3},
\label{eq:condition1}
\ee
which is typically much smaller than the value $\xi_{CR}\sim 10\%$ which is required of SNRs to be the sources of the bulk of Galactic CRs. It follows that in phases in which the SNR accelerates CRs most effectively the growth rate proceeds in a different regime. 

In such regime, that occurs when Eq. \ref{eq:condition} is not fulfilled, the solution of the dispersion relation for $k r_{L,0}\leq 1$, namely for waves that can resonate with protons in the spectrum of accelerated particles ($\gamma\geq \gamma_{min}$) becomes:
\be
\omega_{I} \approx \omega_{R} =  \left[ \frac{\pi}{8} \Omega_{p}^{*} k V_{sh} \frac{n_{CR}(p>p_{res}(k))}{n_{i}} \right]^{1/2}.
\ee
Since $n_{CR}(p>p_{res}(k))\propto p_{res}^{-1}\sim k$, it follows that $\omega\propto k$ for $k r_{L,0}\leq 1$, but the phase velocity of the waves $v_{\phi}=\omega_{R}/k\gg v_{A}$. The fact that the phase velocity of these waves exceeds the Alfv\'en speed may affect the slope of the spectrum of particles accelerated at the shock, as discussed above. 

Even neglecting damping, and requiring that the saturation of the process of magnetic field amplification is solely due to the finite advection time, one can easily show that in both regimes described above at most one can achieve $\delta B/B_{0}\sim 1$, that falls short of solving the problem of reaching the knee by more than one order of magnitude. 

\cite{Bell:2004p737,2005MNRAS.358.181B} noticed that when the condition in Eq. \ref{eq:condition} is violated, namely when 
\be
\xi_{CR} > \frac{\gamma_{min}\Lambda}{3} \left( \frac{v_{A}}{V_{sh}}\right)^{2}\frac{c}{V_{sh}},
\ee
the right hand polarized mode develops a non-resonant branch for $kr_{L,0}>1$ (spatial scales smaller than the Larmor radius of all the particles in the spectrum of accelerated particles), with a growth rate that keeps increasing proportional to $k^{1/2}$ and reaches a maximum for 
\be
k_{M}r_{L,0} = \frac{3\xi_{CR}\gamma_{min}}{\Lambda}\left( \frac{V_{sh}}{v_{A}}\right)^{2}\frac{V_{sh}}{c}>1,
\ee
which is a factor $(k_{M}r_{L,0})^{1/2}$ larger than the growth rate of the resonant mode at $k r_{L,0}=1$. This non-resonant mode has several interesting aspects: first, it is current driven, but the current that is responsible for the appearance of this mode is the return current induced in the background plasma by the CR current. The fact that the return current is made of electrons moving with respect to protons is the physical reason for these modes developing on small scales (electrons in the background plasma have very low energy) and right-hand polarized. Second, the growth of these modes, when they exist, is very fast for high speed shocks, however they cannot resonate with CR particles because their scale is much smaller than the Larmor radius of any particles at the shock. On the other hand, it was shown that the growth of these modes leads to the formation of complex structures: flux tubes form, that appear to be organized on large spatial scales \cite{2012MNRAS.419.2433R} and ions are expelled from these tubes thereby inducing the formation of density perturbations. 

The problem of particle acceleration at SNR shocks in the presence of small scale turbulence generated by the growth of the non-resonant mode was studied numerically in Ref. \cite{2008ApJ.678.255Z}, where maximum energies of the order to $10^{5}$ GeV were found. 

The non-linear development of CR induced magnetic field amplification is likely to be much more complex than illustrated so far. While there is no doubt that the small scale non-resonant instability \cite{Bell:2004p737} is very fast, provided the acceleration efficiency is large enough, the question of what happens to these modes while they grow is complex. MHD simulations \cite{Bell:2004p737}, Particle-in-Cell simulations of this instability \cite{2009ApJ.694.626R} as well as hybrid simulations \cite{2013ApJ...765L..20C} show how the growth of the instability leads to the development of modes on larger spatial scales. 

In addition to a quantitative issue, there is a novel qualitative view of the role of this instability, mainly discussed in Refs. \cite{Schure:2013p3169,2013MNRAS.431.415B}, emphasizing the role of escaping particles as the main responsible for the acceleration process. Particles of a given maximum energy at a given time are able to leave the system because of the absence of waves that they can scatter against. This creates a current of escaping particles which is now responsible for the growth of waves that can inhibit the escape of other particles of the same energy at a later time. 

An estimate of the wavenumber of waves with the maximum growth, $k_{M}$, can be written as
\be
k_{M} B_{0} = \frac{4\pi}{c} j_{CR}.
\ee
In the assumption that the current of escaping particles is quasi-monoenergetic at energy $E$, it can be written as $j_{CR}=n_{CR} \left( \frac{m_{p}c^{2}}{E}\right) e v_{s}$, where $n_{CR}$ is the total number density of CR particles at the shock surface, with energy above $m_{p}c^{2}$. As we discussed above, $k_{M}r_{L}(E) \gg 1$. 

The growth rate of the maximally growing mode can be written in an easy way as
\be
\gamma_{max} = k_{M} v_{A} = \frac{4\pi}{c} \frac{n_{CR} \left( \frac{m_{p}c^{2}}{E}\right) e v_{s}}{\sqrt{4\pi \rho}}.
\ee
As discussed in Refs. \cite{Schure:2013p3169,2013MNRAS.431.415B} the saturation of the instability takes place when $\gamma_{max}\tau\sim 5$, which allows to estimate the maximum energy reachable in different types of supernovae as a function of time. 

Imposing the condition $\gamma_{max} \tau\sim 5$ for the case of a SNR in a uniform background medium of density $\rho$ one easily obtains the maximum energy to be \cite{2014MNRAS.437.2802S,2013MNRAS.431.415B}:
\be
E_{M} = \frac{\xi_{CR}}{10\Lambda} \frac{\sqrt{4\pi \rho}}{c} e R v_{s}^{2},
\ee
where $R$ is the radius of the blast wave. As a numerical value we can assume the radius $R$ to coincide with the beginning of the Sedov phase, when an amount of mass equal to the mass of the ejecta has been processed, $R\sim 2 ~ M_{ej,\odot}^{1/3} n_{ISM}^{1/3}~ pc$, and the interstellar medium density $n_{ISM}$ is normalized to 1 $cm^{-3}$. The shock velocity is normalized by using the total energy of the SN explosion, $(1/2) M_{ej} v_{s}^{2} = E_{SN}$, where $M_{ej}$ is the mass of ejecta. It follows that 
\be
E_{M} = \frac{\xi_{CR} 3^{1/3}}{5 \Lambda} \frac{e}{c} \left( 4\pi \rho \right)^{1/6} E_{SN} M_{ej}^{-2/3} \approx 2\times 10^{5}~\rm GeV,
\ee
where the numerical value has been obtained by assuming that $M_{ej}=1~M_{\odot}$, $n_{ISM}=1~cm^{-3}$ and $E_{SN}=10^{51}$ erg. As discussed in Refs. \cite{2014MNRAS.437.2802S,2013MNRAS.431.415B}, a typical SN exploding in the ISM has a maximum energy that is likely in the range of a few hundred TeV, falling short of the knee by about one order of magnitude. 

The argument presented above can also be generalized to the case of a supernova explosion that takes place in the red supergiant wind of the parent star and leads to \cite{2014MNRAS.437.2802S,2013MNRAS.431.415B}:
\be
E_{M}(R) = \frac{\xi_{CR} e}{5\Lambda c} \sqrt{\frac{\dot M}{v_{W}}} v_{s}^{2}(R),
\label{eq:wind}
\ee
where $\dot M$ and $v_{W}$ are the rate of mass loss and the wind velocity of the progenitor star.

For the case of expansion in the wind, the Sedov phase starts at a distance $R=M_{ej} v_{W} / \dot M$ from the location of the explosion and the shock velocity scales with time as $v_{s} \sim t^{-\frac{1}{m-2}}$, where $m$ is the parameter that describes the shape of the density profile of the ejecta ($\rho_{ej}\propto r^{-m}$). The parameter $m$ is usually assumed to be $m=9$ for core collapse supernovae and $m=7$ for type Ia supernovae. The value $m=9$ leads to $v_{s}\sim t^{-1/7}$, namely the velocity drop is rather slow during the ejecta dominated phase of the expansion, that for typical values $\dot M = 10^{-5} M_{\odot} yr^{-1}$, $v_{W}=10$ km/s and $M_{ej}=1 M_{\odot}$ lasts for about 50 years for ejecta velocity of order $20000$ km/s (corresponding to a total energy $E_{SN}\sim 4\times 10^{51}$ erg). Using these values in Eq. \ref{eq:wind}, we get a maximum energy $E_{M}\approx 2 \times 10^{6}$ GeV, close to the position of the knee. Given the time dependence of the shock velocity, one can envision that the maximum energy may actually grow to larger values at earlier times during the ejecta dominated phase, although the energy contained in this high energy end of the spectrum is bound to be a small fraction of the total energy because of the small mass processed during such stage. It is worth stressing that pushing the parameters to somewhat extreme values ($\xi_{CR}=0.2$, higher mass loss rate, $\dot M=10^{-4} M_{\odot} yr^{-1}$ and $V_{s}=30000$ km/s) one can possibly push the maximum energy up to $E_{M}\sim 2\times 10^{7}$ GeV, although such types of supernovae, if they exist at all, are bound to be much more rare than standard type II supernova events. 

\subsection{Gamma ray observations}

Recent gamma ray observations of SNRs have provided us with a powerful test of the SNR paradigm for the origin of CRs. There is no lack of evidence of CR proton acceleration in SNRs close to molecular clouds (MC), that act as a target for hadronic interactions resulting in pion production. Recently the AGILE \cite{2011ApJ.742L.30G,2010A&A.516L11G,2011ApJ.742L.30G} and Fermi-LAT \cite{2010ApJ.718.348A,Ackermann:2013p3110,2010ApJ.712.459A,2010Sci.327.1103A,2009ApJ.706L.1A} collaborations claimed the detection of the much sought-after pion bump in the gamma ray spectrum. This spectral feature confirms that the bulk of the gamma ray emission in these objects is due to $pp\to \pi^{0} \to 2\gamma$. These cases, besides confirming the existing accelerated hadrons, are very important indicators of CR propagation around the sources \cite{2007Ap&SS.309.365G,2009MNRAS.396.1629G,2008ApJ...689..213R,Nava:2013p3140,2013PhRvD.88b3010G}. On the other hand, not much can be learned from SNRs close to MCs on the acceleration of the bulk of CRs in the Galaxy, since such SNRs are usually not young and the maximum energies are not very high.  

In this perspective, the cases of young individual SNRs are more instructive. The first clear detection of TeV gamma ray emission from a relatively young SNR came from the SNR RXJ1713.7-3946 \cite{2004Natur.432.75A,2006A&A.449.223A,2007A&A.464.235A}, later followed by the detection of the same remnant in the GeV energy range with the Fermi-LAT telescope \cite{2011ApJ.734.28A}. Here I will briefly discuss this case because it is instructive of how the comparison of theoretical predictions with data can drive our understanding of the acceleration environment. 

A discussion of the implications of the TeV data, together with the X-ray data on spectrum and morphology was presented in \cite{Morlino:2009p140}. A hadronic origin of the gamma ray emission would easily account for the bright X-ray rims (requiring a magnetic field of $\sim 160\mu G$), as well as for the gamma ray spectrum.  If electrons were to share the same temperature as protons, the model would predict a powerful thermal X-ray emission, which is not detected. Rather than disproving this possibility, this finding might be the confirmation of the expectation that at fast collisionless shocks electrons fail to reach thermal equilibrium with protons. In fact, the Coulomb collision time scale for this remnant turns out to exceed its age. On the other hand, it was pointed out in Ref. \cite{2010ApJ.712.287E} that even a slow rate of Coulomb scattering would be able to heat electrons to a temperature $\gtrsim 1$ keV, so that oxygen lines would be excited and they would dominate the thermal emission. These lines are not observed, thereby leading to a severe upper limit on the density of gas in the shock region, that would result in a too small pion production. Ref. \cite{2010ApJ.712.287E} concluded that the observed gamma ray emission in this remnant is of leptonic origin. This interpretation appears to be confirmed by Fermi-LAT data, that show a very hard gamma ray spectrum, incompatible with an origin related to pion production and decay. Clearly this does not necessarily mean that CRs are not efficiently accelerated. It simply implies that the gas density in this remnant is, on average, too low for efficient pp scattering to contribute to the gamma ray spectrum. 

However, it should be pointed out that models based on ICS of high energy electrons are not problem free: first, as pointed out in \cite{Morlino:2009p140}, the density of IR light necessary to explain the HESS data as the result of ICS is $\sim 25$ times larger than expected. Second, the ICS interpretation requires a weak magnetic field of order $\sim 10 \mu G$, incompatible with the observed X-ray rims. Finally, recent data on the distribution of atomic and molecular hydrogen around SNR RXJ1713.7-3946 \cite{2012ApJ.746.82F} suggest a rather good spatial correlation between the distribution of this gas and the TeV gamma ray emission, which would be easier to explain if gamma rays were the result of $pp$ scattering. In conclusion, despite the fact that the shape of the spectrum of gamma rays would suggest a leptonic origin, the case of SNR RXJ1713.7-3946 will probably turn out to be one of those cases in which the complexity of the environment around the remnant plays a crucial role in determining the observed spectrum. Future high resolution gamma ray observations, possibly with the Cherenkov telescope array (CTA), will contribute to clarify this situation. 

A somewhat clearer case is that of the Tycho SNR, the leftover of a SN type Ia exploded in a roughly homogeneous ISM, as confirmed by the regular circular shape of the remnant. Tycho is one of the historical SNRs, as it was observed by Tycho Brahe in 1572. The multifrequency spectrum of Tycho extends from the radio band to gamma rays, and a thin X-ray rim is observed all around the remnant. It has been argued that the spectrum of gamma rays observed by Fermi-LAT \cite{2012ApJ744L2G} in the GeV range and by VERITAS \cite{2011ApJ730L20A} in the TeV range can only be compatible with a hadronic origin \cite{Morlino:2012p2243}. The morphology of the X-ray emission, resulting from synchrotron radiation of electrons in the magnetic field at the shock, is consistent with a magnetic field of $\sim 300 \mu G$, which implies a maximum energy of accelerated protons of $\sim 500$ TeV. A hadronic origin of the gamma ray emission has also been claimed by \cite{2013ApJ76314B}, where the steep gamma ray spectrum measured from Tycho is attributed to an environmental effect: the low energy flux would be boosted because the shock is assumed to encounter small dense clumps of material. The flux of high energy gamma rays is not affected because the clumps are assumed to have a small size compared with the scattering length of the protons responsible for the production of TeV gamma rays. In the calculations of \cite{Morlino:2012p2243} the steep spectrum is instead explained as a result of NLDSA in the presence of waves moving with the Alfv\'en velocity calculated in the amplified magnetic field. In this latter case the shape of the spectrum is related, though in a model dependent way, to the strength of the amplified magnetic field, which is the same quantity relevant to determine the X-ray morphology. In the former model the steep spectrum might not be found in another SNR in the same conditions, in the absence of the small scale density perturbations assumed by the authors. 

The case of Tycho is instructive as an illustration of the level of credibility of calculations based on the theory of NLDSA: the different techniques agree fairly well (see \cite{2010MNRAS.407.1773C} for a discussion of this point) as long as only the dynamical reaction of accelerated particles on the shock is included. When magnetic effects are taken into account, the situation becomes more complex: in the calculations based on the semi-analytical description of \cite{Amato:2006p139} the field is estimated from the growth rate and the dynamical reaction of the magnetic field on the shock is taken into account \cite{Caprioli:2008p123,Caprioli:2009p157}. Similar assumptions are adopted in \cite{2008ApJ.688.1084V}, although the technique is profoundly different. Similar considerations hold for \cite{Ptuskin:2010p1025}. On the other hand, \cite{2013ApJ76314B} take the magnetic field as a parameter of the problem, chosen to fit the observations, and its dynamical reaction is not included in the calculations. The magnetic backreaction, as discussed by \cite{Caprioli:2008p123,Caprioli:2009p157} comes into play when the magnetic pressure exceeds the thermal pressure upstream, and leads to a reduction of the compression factor at the subshock, namely less concave spectra. Even softer spectra are obtained if one introduces a recipe for the velocity of the scattering centers \cite{Ptuskin:2010p1025,Caprioli:2010p133,Morlino:2012p2243}. This, yet speculative, effect is not included in any of the other approaches.

Even more pronounced differences arise when environmental effects are included. The case of Tycho is again useful in this respect: the predictions of the standard NLDSA theory would not be able to explain the observed gamma ray spectrum from this SNR. But assuming the existence of {\it ad hoc} density fluctuations, may change the volume integrated gamma ray spectrum as to make it similar to the observed one \cite{2013ApJ76314B}. Space resolved gamma ray observations would help clarify the role of these environmental effects in forging the gamma ray spectrum of a SNR. 

\subsection{DSA in partially ionised media}

The process of particle acceleration at collisionless shocks with velocity $\lesssim 4000$ km/s is deeply affected by the presence of neutral hydrogen in the acceleration region \cite{2012ApJ.755.121B} mainly because of the phenomenon of neutral return flux, which leads to the formation of a neutral induced precursor upstream of the shock, on scales of the order of the pathlength of charge exchange between hydrogen atoms and ions. In typical conditions for SNR shocks in the ISM, this shock modification leads to a steepening of the spectra of accelerated particles at sub TeV energies \cite{2012ApJ.755.121B}. 

As discussed in \S \ref{sec:accel}, the CR induced non-linear effects cause the downstream plasma to be cooler than in the absence of CRs, and to the formation of a CR induced precursor, where the upstream plasma is slowed down (in the shock frame). In the presence of neutral hydrogen in the shock region, both these phenomena reflect in a modification of the Balmer emission \cite{2012p3105,2013ApJ.768.148M,2013A&A...558A..25M}: {\it a)} the broad Balmer line, that carries information on the ion temperature downstream, is expected to be narrower when acceleration is efficient; {\it b)}: the narrow Balmer line, produced by neutrals that suffered charge exchange in the CR induced shock precursor, reflects the temperature of the plasma upstream and is expected to be broader in case of efficient particle acceleration; {\it c)}: an intermediate component of the Balmer line is expected, because of the effect of the neutral return flux \cite{2012p3105}.

These three effects have been recently discussed in the context of a non-linear theory of CR acceleration in partially ionized plasmas developed in Ref. \cite{2013ApJ.768.148M}. The theory has also been used to describe the Balmer emission from SNR 0509-67.5 \cite{2013A&A...557A.142M} and RCW86 \cite{2014A&A...562A.141M}. These cases were used to illustrate how to use observations of the Balmer line to derive estimates of the CR acceleration efficiency in collisionless shocks. 

\section{Transition from Galactic to extragalactic CRs}
\label{sec:trans}

If the maximum energy of protons accelerated in SNRs is as high as the knee, $\sim 3000$ TeV, then the superposition of spectra of elements with higher mass can fit the structure of the knee, as discussed in \cite{Blasi:2012p2051}, though the all-particle spectrum can be shown to start diving down already at energies around $\sim 10^{16}$ eV. This fact was also used by \cite{Hillas:2005p171} to infer the existence of an additional population of sources able to reach even larger maximum energies. In any case, the Galactic spectrum is expected to end with a heavy mass composition at energies around a few $10^{17}$ eV. 

The end of the Galactic CR spectrum and the transition to extragalactic CRs become then strictly connected with how the extragalactic CR component starts and with which mass composition. While at $\sim 10^{18}$ eV all largest experiments, Auger \cite{Abraham:2010yv}, TA
\cite{Jui:2011vm,Tsunesada:2011mp} and HiRes
\cite{Abbasi:2002ta,Abbasi:2007sv}, agree that the flux of ultra high energy cosmic rays (UHECRs) is dominated by a light component, at higher energies the situation is somewhat less well defined. The Auger observatory measured the mean depth of shower maximum ($X_{max}(E)$) and its fluctuations ($\sigma(E)$) as functions of the primary energy $E$ and found evidence for a gradual transition to a heavier mass composition with increasing energy. This result, by itself, would rule out the so-called dip model \cite{Berezinsky:2002nc,Aloisio:2006wv} for the transition from Galactic to extragalactic CRs. 

The issue was studied in detail in \cite{2013arXiv1312.7459A,2014APh....54...48T} where the authors reach the conclusion that a satisfactory fit to the Auger spectrum and mass composition can only be achieved by assuming a hard injection spectrum with slope $\sim 1-1.6$, at odds with the predictions of the most common acceleration mechanisms, that lead to steeper spectra.

The most striking consequence of the hardness of the injection spectra is however that only the highest energy part of the spectrum can be fitted ($E\gtrsim 5\times 10^{18}$ eV) leaving another gap in the spectrum at lower energies: the conclusion as derived in \cite{2013arXiv1312.7459A} is that an additional component is needed to fit the all-particle spectrum, and such a component must be mostly light in order to accomodate the light composition measured by HiRes \cite{Abbasi:2002ta,Abbasi:2007sv}, TA \cite{Jui:2011vm,Tsunesada:2011mp} and Auger \cite{Abraham:2010yv} at $10^{18}$ eV, and must be of extragalactic origin in order to avoid contradiction with the Auger measured anisotropy \cite{2013ApJ...762L..13P}. Moreover, this extra component is required to have a steep spectrum with slope $\sim 2.7-2.8$ and a relatively low maximum energy, $\sim (5-10)\times 10^{18}$ eV, otherwise the fit to $X_{max}(E)$ and $\sigma(E)$ at high energies would be negatively affected. One should appreciate the change of paradigm that resulted from Auger data: while a decade ago the main theoretical challenge was that of finding sources (or acceleration mechanisms) able to produce protons up to energies of $\gtrsim 10^{20}$, the present Auger mass composition forces us to have protons up to energies of $\lesssim 10^{19}$ eV, while particles at the highest energies are expected to be heavier nuclei. 

The recent KASCADE-Grande measurement \cite{Apel:2013ura} of the spectrum and chemical composition in the energy region between $10^{16}$ eV and $10^{18}$ eV has shed an interesting light on the transition region: on one hand these measurements show evidence of a feature in the light component in the form of an ankle-like structure at $\sim 10^{17}$ eV. This spectral feature was interpreted by the collaboration as the possible signature of a transition between a Galactic component (extending to $\sim 10^{17}$ eV) and an extragalactic light component (with slope $2.79\pm 0.08$). On the other hand, the same measurements show that the spectrum of the heavy CR component has a knee at $\sim 10^{17}$ eV, where it steepens from a slope $2.95\pm 0.05$ to a slope $3.24\pm 0.08$). In Ref. \cite{2013arXiv1312.7459A} it was shown that the spectrum of the heavy component cannot be easily interpreted as standard Galactic iron in that it requires sources able to accelerate iron to $\sim 10^{18}$ eV.  

The transition region may be rather complicated, with Galactic CR sources contributing heavy ions up to $\sim 1$ EeV and multiple components of extragalactic CRs, with protons and nuclei injected with different injection spectra. Though not appealing, this complexity may be the way in which Nature realizes the transition from galactic to extragalactic CRs, and it has its own interesting aspects: in fact it has the noticeable implication that, contrary to the simplest intuition, there may be Galactic sources able to accelerate CRs to a rigidity of $\sim (3-4)\times 10^{16}$ V (corresponding to iron nuclei at $\sim 10^{18}$ eV), about an order of magnitude above the rigidity of the knee. This exacerbates the severe difficulties faced by the theory of particle acceleration in SNRs, and might require a different class of sources altogether.

As discussed earlier in this paper, the recent ARGO-YBJ and YAC1+Tibet array data find that the knee in the light component appears at $\sim 700 $ TeV, quite below the knee in the all-particle spectrum, that ARGO confirms being at $\sim 3000$ TeV. This finding, in obvious friction with previous results by the KASCADE experiment, makes the issue of the transition to extragalactic CRs even more problematic and does not relax the requirements for acceleration, since Galactic sources able to accelerate CRs to very high energies are still needed in order to fit the all-particle spectrum.

Better measurements of the spectrum and mass composition in the energy region between $10^{15}$ eV and $10^{18}$ eV, and a careful understanding of the systematic uncertainties and the dependence of the results upon Monte Carlo codes of the shower development are mandatory if we want to understand the end of the Galactic CR spectrum and the transition to the UHECR domain \cite{2014CRPhy..15..329B}.

\section{Summary}
\label{sec:summary}

The connection between the bulk of CRs and SNRs has been and is central to the search for the origin of CRs. Despite many steps forward, the issue of reaching the knee through DSA remains unsolved or, at least, unclear. On one hand, it is clear that plasma instabilities in the shock region must play an important role in the acceleration process, as shown by the observation of thin, non-thermal X-ray filaments in virtually all young SNRs. On the other hand, different flavors of the CR induced streaming instability seem to be barely sufficient to reach $E_{max}\sim 3Z\times 10^{15}$ eV, and only for rather energetic supernovae exploding in the red giant wind of the progenitor star. If SNRs are responsible for CR protons up to $E_{max,p}\sim 3\times 10^{15}$ eV, then the knee should consist of a gradual transition to heavier elements and Galactic CRs should somehow terminate at $E\sim 10^{17}$ eV with an iron dominated composition. The doubts cast by the recent results of YAC1+Tibet and ARGO-YBJ on the position of the knee in the light CR component do not help clarifying the situation. 

Similar experimental ambiguities at lower energies (for instance the discrepancy between PAMELA and AMS-02 on the spectrum of protons and helium nuclei) are hindering our ability to fully understand the origin of CRs. 

One of the priorities in future research in this field should be the accurate measurement of the spectra and mass composition of CRs throughout the knee region. This piece of information can provide precious hints not only on CR acceleration and propagation, but also on the issue of the transition to extragalactic CRs, that represents one of the hottest topics that are currently being debated. 

\section*{Ackowledgements}
This research work was partially funded through Grant PRIN-INAF 2013.



%% References

%\nocite{*}
\bibliographystyle{elsarticle-num}
\bibliography{biblio1}






% \end{thebibliography}

\end{document}